\documentclass[floats,floatfix,showpacs,amssymb,prd,twocolumn,superscriptaddress,nofootinbib,nolongbibliography,reprint]{revtex4-1}

\usepackage{amssymb,amsmath,verbatim,mathtools,needspace,enumitem,etoolbox,graphicx,physics,microtype,afterpage,xspace,tabularx,lmodern,multirow}
\usepackage{gensymb}
\usepackage[normalem]{ulem}
\usepackage[dvipsnames, usenames]{xcolor}
\definecolor{linkcolor}{rgb}{0.0,0.3,0.5}
\usepackage[unicode, colorlinks=true, linkcolor=linkcolor, citecolor=linkcolor, filecolor=linkcolor, urlcolor=linkcolor, linktocpage, breaklinks]{hyperref}
\usepackage[all]{hypcap}
\usepackage{subfigure}
\usepackage[T1]{fontenc}
\usepackage[utf8]{inputenc}
\usepackage[usenames,dvipsnames]{xcolor}
\usepackage{enumerate}

\hypersetup{colorlinks=true,citecolor=romared,linkcolor=romared,urlcolor=romared}

\definecolor{romared}{RGB}{142,0,28}

\newcommand{\be}{\begin{equation}}
\newcommand{\ee}{\end{equation}}

\def\be{\begin{equation}}
\def\ee{\end{equation}}
\newcommand{\beq}{\begin{eqnarray}}
\newcommand{\eeq}{\end{eqnarray}}

\usepackage{aas_macros}
\usepackage{makecell}
\usepackage{soul}
\usepackage{amssymb}
\usepackage{longtable}

\usepackage{lipsum}

\newcolumntype{Y}{>{\centering\arraybackslash}X}

\begin{document}

\title{On the physical significance of black hole quasinormal mode spectra instability}

\begin{abstract}
It has been shown, via specific examples and a pseudospectrum analysis, that the black hole quasinormal spectra are unstable. The implication of such a result for gravitational-wave physics and of our understanding of black holes is, still, unclear. The purpose of this work is twofold: (i) we show that some of the setups leading to instabilities are unphysical and triggered by exotic matter or extreme spacetimes; (ii) nevertheless, we also show simple examples of compelling physical scenarios leading to spectral instabilities. Our results highlight the importance of understanding the overtone content of time-domain waveforms, and their detectability.
\end{abstract}

\author{Vitor Cardoso} 
\affiliation{Niels Bohr International Academy, Niels Bohr Institute, Blegdamsvej 17, 2100 Copenhagen, Denmark}
\affiliation{CENTRA, Departamento de F\'{\i}sica, Instituto Superior T\'ecnico -- IST, Universidade de Lisboa -- UL, Avenida Rovisco Pais 1, 1049-001 Lisboa, Portugal}

\author{Shilpa Kastha} 
\affiliation{Niels Bohr International Academy, Niels Bohr Institute, Blegdamsvej 17, 2100 Copenhagen, Denmark}

\author{Rodrigo Panosso Macedo} 
\affiliation{Niels Bohr International Academy, Niels Bohr Institute, Blegdamsvej 17, 2100 Copenhagen, Denmark}

\maketitle

\section{Introduction}

In a binary black hole (BH) coalescence process, as the two BHs merge they leave behind a ``distorted'' remnant, emitting gravitational waves (GWs) and evolving towards stationarity via a {\it ringdown} phase. During this phase, the GW signal is simply described by a superposition of exponentially damped sinusoids termed quasi-normal modes (QNMs). Thus, BHs relax as most systems do, by oscillating in a set of characteristic frequencies~\cite{Kokkotas:1999bd,Berti:2009kk,Konoplya:2011qq,Barausse:2014tra}.

In a way similar to the analysis of the Hydrogen atom, the response of (non-spinning) BH spacetimes is decomposed in spherical harmonics $Y_{\ell m}$, and each QNM carries, in addition to the polar and azimuthal indices $(\ell, m)$ an overtone index $n=0,1,2...$. The intrinsic dissipative character of BH spacetimes implies that the modes are not stationary (energy is being radiated to large distances and towards the horizon), and characteristic frequencies are complex numbers. For each $(\ell, m)$, the dominant mode is the one with lowest imaginary component of the characteristic frequency, and assigned the $n=0$ label, the fundamental mode.
In vacuum General Relativity (GR), the characteristic QNM frequencies of a single mode are enough to  characterize the underlying geometry of the remnant. BH spectroscopy is the science of extracting information about BH spacetimes and the underlying gravitational description from the relaxation of BHs. Detection of these modes from a binary BH merger process provides information about the object itself, its surrounding environment, and even potential deviations from General Relativity~\cite{Kokkotas:1999bd,Berti:2009kk,Konoplya:2011qq}.

The first detection of a fundamental QNM is established for the first GW event, GW150914~\cite{LIGOScientific:2016aoc}, where the fundamental quadrupolar mode $(\ell,m,n) = (2,2,0)$ mode is reported to be consistent with General Relativity predictions~\cite{LIGOScientific:2016lio}. The ringdown of the BH remnant has since been reported for a number of other events (see in particular tables VIII of Ref.~\cite{LIGOScientific:2020tif} and XI of Ref.~\cite{LIGOScientific:2021sio}). Detection of higher overtones will require louder signals and careful analysis~\cite{Cotesta:2022pci,Carullo:2023gtf,Baibhav:2023clw,Nee:2023osy,Zhu:2023mzv}. Tentative evidence for additional modes $(\ell,m,n) = (3,3,0)$ was reported for event GW190521~\cite{Capano:2021etf,Capano:2022zqm,Forteza:2022tgq,Abedi:2023kot} but is the subject of debate, given systematic uncertainties such as precession and eccentricity. 
The impact of nonlinearities has recently been discussed~\cite{Cheung:2022rbm, Mitman:2022qdl, Zlochower:2003yh,Sberna:2021eui,Baibhav:2023clw,Redondo-Yuste:2023seq}. In other words, BH spectroscopy is now blossoming into a vibrant field, and upcoming years will see precision tests of General Relativity from the relaxation of BHs~\cite{Dreyer:2003bv,Berti:2005ys,Berti:2007zu,Berti:2016lat}.

The BH spectroscopy program anchors on linearized calculations of the vacuum BH spectrum in GR~\cite{Kokkotas:1999bd,Berti:2009kk,Konoplya:2011qq,Barausse:2014tra,Dreyer:2003bv,Berti:2005ys}, and the paradigm implicitly assumes a reasonable physical robustness against ``small'' changes in the system. 
In colloquial terms, one does not expect a ``flea'' far away from the BH to perturb significantly its spectrum. Surprisingly, such a spectral instability has been observed in a variety of circumstances. 

Indeed, in addition to his pioneering work on the description of ringdown dynamics~\cite{VishuNature70}, Vishveshwara also explicitly pointed out that small changes in the BH potential may lead to a destabilization of the QNM spectra~\cite{Aguirregabiria:1996zy}\footnote{When recollecting his journey along the BH trail~\cite{Vishveshwara:1996jgz}, Vishveshwara mentions that ``we have studied the sensitivity of the QNMs to scattering potentials. The motivation is to understand how any perturbing influence, such as another gravitating source, that might alter the effective potential would thereby affect the QNMs. Interestingly, we find that the fundamental mode is, in general, insensitive to small changes in the potential, whereas the higher modes could alter drastically. The fundamental mode would therefore carry the imprint of the BH, while higher modes might indicate the nature of the perturbing source."}. The sensitivity of the QNMs under small modifications of the potential has been somewhat rediscovered recurrently over the past decades~\cite{Nollert:1996rf,Nollert:1998ys,Leung:1997was,Motl:2003cd,Andersson:2003fh,Natario:2004jd,Berti:2009kk,Barausse:2014tra,Jaramillo:2021tmt,Cheung:2022rbm,Konoplya:2022hll,Konoplya:2022pbc}, and it has recently been incorporated into a more formal framework~\cite{Jaramillo:2021tmt,Jaramillo:2020tuu}.
Unfortunately, with notable exceptions discussed below, spectral instabilities have been discussed in the context of ad hoc perturbations to the effective potential, the physical significance, if any, of which is not fully understood. The purpose of the present work is to shed {\it some} light on these issues.

Broadly speaking, one can identify two major classes of spectral instabilities, closely related to the ``infra-red'' and ``ultra-violet'' effects reported in Ref.~\cite{Jaramillo:2020tuu} (but we are aware of the dangers of any classification such as this one):

\noindent {\bf Fundamental mode instabilities}, which are associated with the introduction of a different scale in the problem. These include shells of matter for example, modelled as ``potential bumps'' far away from the BH~\cite{Barausse:2014tra,Cheung:2022rbm,Berti:2022xfj}, or abrupt cuts in the large range of the potential~\cite{Nollert:1996rf, Nollert:1998ys}. As we argue below, the new scale means that matter is present either asymptotically far or asymptotically close to the BH horizon, leading in either case to low-frequency echoes. These do indeed de-stabilize the spectrum, by introducing a longer lived second family of modes.

The time-domain response is only affected after the prompt ringdown emission, which is governed by the vacuum QNMs.

\noindent {\bf Overtone instabilities}, caused by high-momentum fluctuations or seemingly near-horizon modifications to the geometry~\cite{Leung:1997was, Barausse:2014tra,Jaramillo:2020tuu, Jaramillo:2021tmt, Cheung:2022rbm,Konoplya:2022hll,Konoplya:2022pbc}. These affect the larger overtones, and potentially also affect promptly the time-domain signal\cite{Jaramillo:2021tmt}, but a systematic understanding is lacking.

We will show how the existing body of work relates to previous known results, how the spectral instability can arise in physically interesting setups, and how some of the scenarios considered thus far require exotic matter or extreme spacetimes.

\section{Setup}
To include in our analysis the relevant results reported in the literature, we consider a general, spherically symmetric spacetime
\be
\label{eq:metric}
ds^2=-a(r) dt^2+\dfrac{dr^2}{b(r)}+r^2d\Omega^2\,,
\ee
The mass function $m(r)$ is defined from the metric via
\be
b(r) = 1-\dfrac{2m(r)}{r}\,.
\ee
The energy density associated to the mass function is
\be
\label{eq:density}
\rho(r) = \dfrac{m'(r)}{4 \pi r^2}.
\ee

The dynamical equations governing massless fields follows from the field equations and equation of state of matter. The details of the particular equation of state of matter are unnecessary. Indeed, all analyses in the literature focused exclusively on scalar fields and on only one sector of gravitational perturbations, for which the the massless field does not couple to the background matter (assuming it is isotropic and dissipationless).

Both scalar and gravitational axial fluctuations can be expanded in harmonics of index $\ell$, and are governed by a master wavefunction $\Psi=\Psi(t,r)$ which obeys the second order partial differential equation,
\beq
\dfrac{\partial^2\Psi}{\partial r_*^2}-\dfrac{\partial^2\Psi}{\partial t^2}-V\Psi=0\,,\label{eq:WaveEq_universal}
\eeq
where the tortoise coordinate is defined by $\mathrm{d}r/\mathrm{d}r_*=\sqrt{ab}$.

For minimally coupled scalars $\Phi=\Psi/rY_{\ell m}$, one finds that the field is governed by~\cite{Berti:2009kk}
\beq
V=a\left(\dfrac{\ell(\ell+1)}{r^2}+\dfrac{(ab)'}{2ar}\right)\,.\label{eq:WaveEq_scalar}
\eeq

In spherical symmetry, gravitational fluctuations can be decomposed into two sectors (axial or odd, and polar or even). The axial sector, for isotropic dissipationless fluids does not couple to the matter and obeys the following~\cite{Cardoso:2021wlq,Redondo-Yuste:2023ipg}\footnote{Note that Chandrasekhar and Ferrari~\cite{Chandrasekhar:1991fi} and
Kokkotas find ($M$ is total mass of star)~\cite{Kokkotas:1999bd}
\be
V=a\left(\dfrac{\ell(\ell+1)}{r^2}-\dfrac{6M}{r^3}+\rho-P\right)\,.
\ee
There are $4\pi$ factors difference between the two equations but these are due to the different convention in the field equations. Equation \eqref{eq:AxialWaveEq_Gen} uses $G_{\mu\nu}=8\pi T_{\mu\nu}$, whereas Chandra uses $G_{\mu\nu}=2 T_{\mu\nu}$.
}
\be
V=a\left(\dfrac{\ell(\ell+1)}{r^2}-\dfrac{6m}{r^3}+\dfrac{m'}{r^2}\right)\,.\label{eq:AxialWaveEq_Gen}
\ee
%

\section{Spectral instability of the fundamental mode: a new scale}

\subsection{Double-bump potentials sourcing instabilities of the fundamental mode: 
is it really a flea? \label{sec:Flea}}

Spectral stability concerns the robustness of the QNM spectrum against slight deviations of the underlying system~\cite{Aguirregabiria:1996zy,Nollert:1996rf,Barausse:2014tra,Jaramillo:2020tuu,Jaramillo:2021tmt}. Some studies have focused on changes to the effective potential governing wave propagation, as~\cite{Aguirregabiria:1996zy,Cheung:2021bol,Berti:2022xfj}
\be
V_{\epsilon}=V+\epsilon V_{\rm bump}\,,
\ee
with $V$ the ``unperturbed'' potential describing propagation in vacuum Schwarzschild BH spacetime (i.e. setting $m'=0,\,m=M_{\rm BH}$ in \eqref{eq:WaveEq_scalar}-\eqref{eq:AxialWaveEq_Gen}). The focus was then on ``bumpy'' potentials $V_{\rm bump}$, such that $V_{\epsilon}$ admits a local maximum besides the global maximum of $V$.

Notice that the system will have distinct scales when the characteristic distance $L$ of the bump away from the BH is much larger than any other scale in the problem, in this case the BH mass. In this circumstance, there is a cavity formed by the two potential peaks which supports a mode of frequency $\sim 1/L$, as long as it is trapped, $1/L^2\ll \epsilon V_{\rm bump}$. In these circumstances, the BH fundamental QNM is de-stabilized~\cite{Barausse:2014tra,Cheung:2021bol,Berti:2022xfj}: for any arbitrarily small $\epsilon V_{\rm bump}$ there is a scale $L$ at which an extremely low frequency mode appears, which dominates over the BH light ring mode~\cite{Cardoso:2016rao,Cardoso:2019rvt} and which is never ``close'' to it.

We have been discussing features asymptotically far from the BH, but similar phenomenology occurs if new scales show up close to the horizon, under the form of matter of simply new boundary conditions~\cite{Cardoso:2016rao,Cardoso:2019rvt}. In all cases, the time-domain response is smooth: the prompt response corresponds to a vacuum BH signal, but at late times the double-bump cavity filters out the high-frequency component, giving rise to a sequence of ``echoes''~\cite{Cardoso:2016rao,Cardoso:2016oxy,Cardoso:2019rvt}.

An open question is what physics gives rise to such a potential. Is it really a ``small'' spacetime perturbation? To establish an equivalence between a ``bumpy'' potential and the potential \eqref{eq:WaveEq_scalar} arising from a generic spherically symmetric spacetime, one can take the eikonal limit $\ell\to \infty$. In this limit, potential \eqref{eq:WaveEq_scalar} asymptotes to 
\be
V=a\dfrac{\ell(\ell+1)}{r^2},\,\ell \to \infty\,.
\ee
The requirement that $V$ has extrema corresponds to the requirement that $2a=ra'$ at some point in the exterior of the horizon. But this requirement is equivalent to the existence of a light ring at that location~\cite{Cardoso:2008bp}. Thus, the spacetime would have {\it two} light rings, and one cannot consider $\epsilon V_{\rm bump}$ to be a small perturbation by any means.

The above reasoning assumes that $(ab)'/r$ decays faster than $1/r^2$, which it should for regularity reasons for all cases. The only exception concerns massive scalar fields of mass $\mu$, which indeed can be looked as a special case of all the above. For such fields the true effective potential is $V+a\mu^2$, we discuss it below.

\subsection{Massive fields: a true fundamental-mode instability}
A very clear example of spectral instabilities concerns massive fields. What we will now discuss affects both massive tensor, vector and scalar fields, but for concreteness we focus on scalars, whose dynamics are governed by the Klein-Gordon equation,
\be
\Box \Phi=\mu^2\Phi\,.
\ee
The mass parameter $\mu$ is related to the physical boson mass $m_B$ via $m_B=\mu\hbar$, in our units.

We focus for simplicity on a Schwarzschild background. For massless fields, $\mu=0$, the fundamental $\ell=0,1$ mode are given by $M\omega= 0.1104- i0.1049, 0.2929 - i0.0977$, respectively. 

For very small $\mu$ (the regime most interesting to us here), two things happen. First, a {\it new} family of modes, so-called quasi-bound states, appears~\cite{Detweiler:1980uk,Cardoso:2005vk,Dolan:2007mj,Brito:2015oca} which in the small $\mu$ limit is well described by~\cite{Pani:2012bp}
\beq
M\omega&=&M\mu-\frac{(M\mu)^2}{2(\ell+n+1)^2}-iM\omega_I\,.
\eeq
For $\ell=1$ for example, $M\omega_I=(M\mu)^{10}/12$.
This is a true spectral instability: for any arbitrarily small
$\mu$ the fundamental mode is changed by a large amount. For an {\it arbitrarily small} mass $\mu$ there are new modes of arbitrarily small frequency. In other words, we argue that among all ``bumpy potentials'' the only physically relevant concerns endowing a small mass to interaction carriers.

The second interesting property is that the massless family (the usual QNM spectrum when $\mu=0$) is altered~\cite{Simone:1991wn,Cardoso:2005vk,Dolan:2007mj}. As an example, the fundamental mode $n=0$ mode changes as
\beq
\frac{\omega^\ell_\mu-\omega^\ell_0}{\mu^2}&=&0.225+ i0.823\,,\qquad \ell=0\,,\\
&=&0.447 + i0.268\,,\qquad \ell=1\,.
\eeq
For small $\mu$ the larger the overtone the smaller the deviation from the $\mu=0$ mode. 

As discussed at length elsewhere, the most immediate effect of such a de-stabilization are drastic changes to the late-time behavior~\cite{Barausse:2014tra,Berti:2022xfj,Cardoso:2016rao,Cardoso:2019rvt}. The impact on prompt ringdown is poorly studied, but it is not our focus here.

\subsection{A holistic view on spectral instability of the fundamental mode: soft changes and couplings to matter~\label{sec:couplings}}
\begin{figure}[t!]
\includegraphics[width=\linewidth]{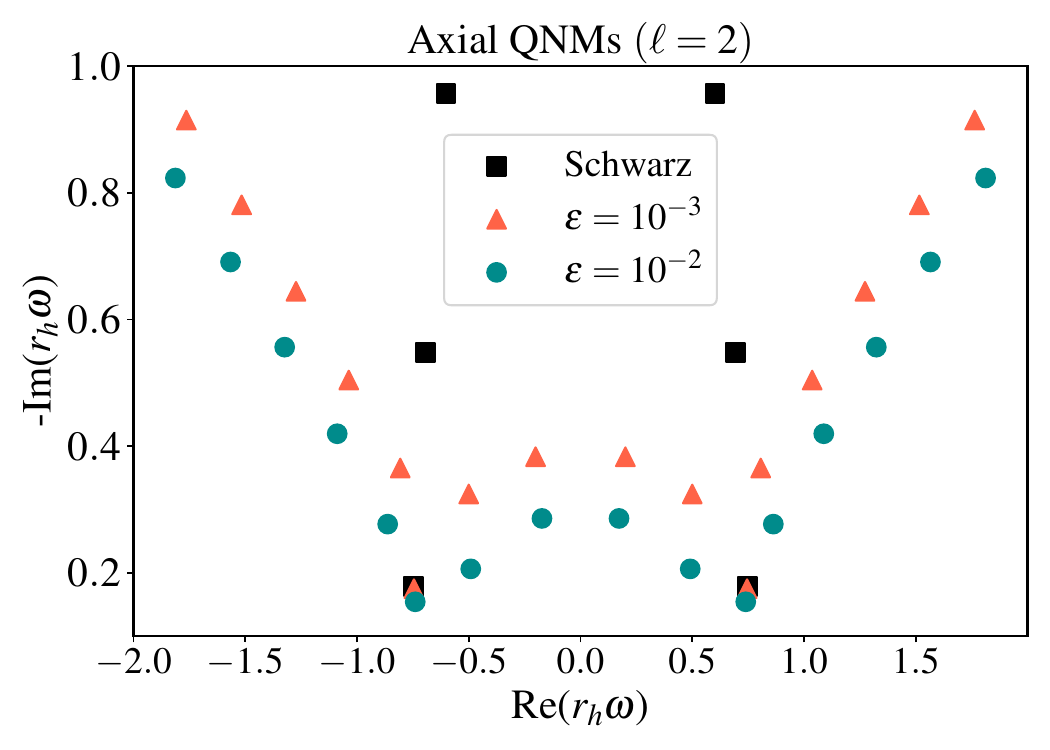}
\caption{Axial QNMs (angular mode $\ell=2$) for potential with a P\"oschl-Teller secondary bump~\cite{Cheung:2021bol} located at $L = 10 r_h$. Black squares show the first three modes of a pure vacuum Schwarzschild spacetime. Triangles and circles are results when a P\"oschl-Teller bump of height $\epsilon=10^{-2}, 10^{-3}$ is introduced. New modes appear, some of low frequency. Higher overtones are exponentially sensitive in overtone number $n$ and scale $L$, see main text.}
\label{fig:QNM_DoubleBump}
\end{figure}
The massive scalar case just discussed is a proxy for more general setups: whenever the effective potential governing propagation is changed, the spectrum changes. If the changes are ``soft'' (induced by a parametrically small quantity) but there are different scales in the problem, then as a rule new families of modes set in, and large scales control low frequency modes. For very small perturbations with scale separation, the vacuum family continues to exist, with slight induced corrections. In addition to the massive scalar field discussed above, these features have been observed for simple two-barrier toy models~\cite{Barausse:2014tra} or for asymptotically de Sitter spacetimes
~\cite{Cardoso:2017soq}. 

Figure~\ref{fig:QNM_DoubleBump} summarizes what happens when a ``soft'' change in the potential is introduced. We take the effective potential for axial gravitational modes and modify it by adding a P\"oschl-Teller secondary bump~\cite{Cheung:2021bol}.
The three lowest modes of the pure vacuum potential have frequencies~\cite{Berti:2009kk}
\beq
r_h\omega/2&=&0.3737-i0.08896\,,\quad n=0\,,\nonumber\\
&=&0.3467-i0.2739\,,\quad\, \,n=1\,,\nonumber\\
&=&0.3010-i0.4783\,,\quad \,\, n=2\,,\nonumber
\eeq
shown as black squares in the figure. It is clear that a small bump introduces new modes in the problem, some of them of lower frequency, a clear sign of instability of the spectrum. Nevertheless the fundamental mode of the vacuum family is only slightly disturbed. This can be understood from a perturbation theory point of view. The introduction of a small disturbing potential affects the modes as~\cite{Leung:1999iq,Barausse:2014tra}
\be
\delta \omega \sim \epsilon \bra{\psi} V_{\rm bump}\ket{\psi}\sim \epsilon e^{2L\omega_I}\,,
\ee
with $L$ the scale at which the bump is located, and $\omega_I$ the imaginary component of the frequency of the mode in question. Thus, to be in the perturbative regime, one has to require that $\epsilon \ll e^{-2L\omega_I}$. 
But if $L= 10 r_h$, $e^{-2L\omega_I}=0.17,4\times 10^{-3}, 7\times 10^{-5}$ for $n=0,1,2$. This explains why overtones are exponentially sensitive to any change in the potential.
We have explicitly verified this exponential sensitivity for the double barrier potential in Ref.~\cite{Barausse:2014tra}.

On the other hand, if there are ``hard'' changes, for instance, an abrupt change of boundary conditions, then as a rule the prompt ringdown, vacuum BH mode is still dominant in the time response, but it does not feature in the spectrum~\cite{Cardoso:2015fga,Cardoso:2016rao,Cardoso:2019rvt}.

From the discussion of Section~\ref{sec:Flea}, one might be tempted to incorrectly conclude that instability of the fundamental mode occurs only in highly contrived, non-physical situations. Asymptotically de Sitter spacetimes are a clear example where the instability is triggered, by the introduction of a new scale. The small cosmological $\Lambda$ constant of our universe adds a new family of low-frequency $\omega \propto \sqrt{\Lambda}$ modes to the BH spectrum~\cite{Cardoso:2017soq,Jansen:2017oag}. 

Most importantly, the majority of physically relevant setups are not included in a simple, decoupled second-order partial differential equation of the form~\eqref{eq:WaveEq_universal}. Indeed, a wide class of problems couple gravitational degrees of freedom to matter modes~\cite{Cardoso:2022whc}. Since matter moves at subliminal speeds, their characteristic frequencies are also lower. The universal coupling of gravity to matter then produces a spectral instability, no matter how low the matter density~\cite{Cardoso:2022whc}. Because of the difference of scales associated to the instability of the fundamental mode, we expect that the prompt ringdown remains essentially unaffected for this class of perturbations.

\section{Spectral instability of overtones}
We now focus on the second class of spectral instabilities, affecting BH overtones. As we show below, some of the setups used to argue for this type of instability are unphysical. Nevertheless, we show that the mechanism itself was known for some time, and can arise in physically relevant situations. With a few notable exceptions~\cite{Jaramillo:2021tmt}, the impact of such an instability in time-domain signals is largely unexplored, and understanding its detectability in more realistic scenarios requires further work.

\subsection{Charged BHs: a true instability}
The existence of a spectral instability should be no surprise, since it was indirectly known in a rather straightforward setup for at least two decades in BH physics~\cite{Motl:2003cd,Andersson:2003fh,Natario:2004jd,Berti:2009kk,Daghigh:2024wcl}. Consider the asymptotic spectrum of non-spinning, neutral BHs, in the large overtone limit, and for any massless field
\be
M\omega \sim \frac{\log 3-i(2n+1)\pi}{8\pi} + {\cal O}(n^{-1/2})\,,\quad Q=0\,,
\ee
a behavior which attracted considerable attention in relation to possible BH area quantization~\cite{Berti:2009kk}. Frequencies are equally spaced in their imaginary component, whereas real part asymptotes to a constant $\log 3/(8\pi M)$. 

On the other hand, an analysis of the asymptotic behavior of BHs charged with an arbitrary charge $Q$ yields the condition
\beq
&&e^{\omega/T} +2+3e^{Q^4\omega/(Tr_+^4 )} =0\,,\quad Q\neq 0\,,\\
&&r_+\equiv M+\sqrt{M^2-Q^2}\,,\quad T\equiv \frac{\sqrt{M^2-Q^2}}{2\pi r_+^2}\,.
\eeq
In the limit that $Q\to 0$, one finds then 
\be
M\omega \sim \frac{\log 5-i(2n+1)\pi}{8\pi} + {\cal O}(n^{-1/2})\,,\quad Q\to 0\,,
\ee

As a consequence for any arbitrarily small charge $Q$ the BH spectrum differs by ${\cal O}(Q^0)$ from the spectrum of a Schwarzschild BH\footnote{\label{fn:mantra}Hence also the mantra that the theory of the limit need not be the limit of the theory. A common, if trivial, example concerns the function $n\epsilon/(1+n\epsilon)$. At $\epsilon=0$ this yields zero at any $n$, even though the large $n$ limit is unity.}. This is a clear example of a spectral instability in the large overtone regime.

\subsection{High-momentum fluctuations}
%
Recently, another class of fluctuations leading to large-overtone spectral instability has been studied, and concerns a modification to the BH (of mass $M_{\rm BH}$) potential in the form~\cite{Jaramillo:2020tuu}
\be
\label{eq:PertPot}
V_\epsilon= V+\dfrac{a}{r^2}\delta V\,,
\ee
with $|\delta V|\sim \epsilon$ a small perturbation. 

Comparing eqs.~\eqref{eq:AxialWaveEq_Gen} and \eqref{eq:PertPot}, it is straightforward to establish a relation between the mass aspect $m(r)$ and the potential modification $\delta V(r)$ via
\be
\label{eq:mass_deltaV_r}
\dfrac{6 m(r)}{r} - m'(r) = \dfrac{6M_{\rm BH}}{r} - \delta V(r)\,.
\ee
A careful study of the asymptotic limit $\displaystyle \lim_{r\rightarrow \infty} m(r)$ reveals that the condition
\be 
\label{eq:RegCond_PertPot}
\lim_{r\rightarrow \infty}\delta V(r) = 0\,,
\ee
is necessary for a regular mass function.

The suggested models capturing high-momentum fluctuations employ $\delta V=\epsilon \cos\left(2\pi k \, r_h/r\right)$~\cite{Jaramillo:2020tuu}, with $k\in {\mathbb N}$. This choice, however, does not satisfy the regularity condition \eqref{eq:RegCond_PertPot}. 
However, the perturbing potential
\be
\label{eq:deltaV_sin}
\delta V(r) = \epsilon \sin\left(2\pi k \, \dfrac{r_h}{r} \right)\,,
\ee
satisfies the regularity condition~\eqref{eq:RegCond_PertPot} and is as well motivated as the original. We verified that Eq.~\eqref{eq:deltaV_sin} yields the same qualitative behavior for the QNM instability~\cite{Jaramillo:2021tmt}.

\begin{figure}[t!]
\includegraphics[width=\linewidth]{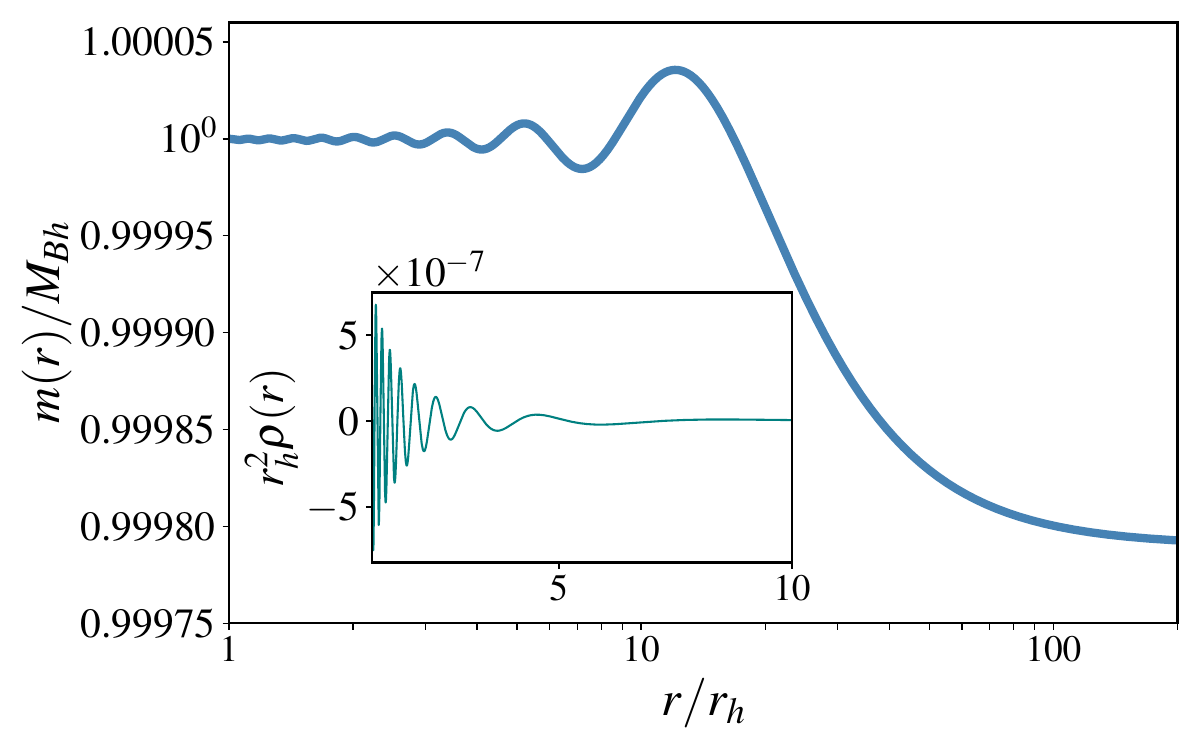}
\caption{Physical quantities associated to the potential $\delta V = \epsilon \sin(2 \pi k \, r_h/r)$ which leads to spectral instabilities. Mass $m(r)$ and density $\rho(r)$ profiles are well defined in the BH exterior region $r\in[r_h,\infty)$. But $\rho(r)$ assumes negative values, which prevents models with a single wavenumber $k$ from being realistic at a classic level. Here, $\epsilon=10^{-5}$, $k=10$.
}
\label{fig:PertMasFunc_DertPer}
\end{figure}

With this choice, eq.~\eqref{eq:mass_deltaV_r} is straightforwardly solved, with integration constant fixed to ensure $m(r)$'s regularity as $r\rightarrow \infty$. The corresponding energy density \eqref{eq:density} reads
\beq
&& r_h^2 \, \rho = -\dfrac{9 r^3 \epsilon }{8 \pi ^6 k^5 r_h^3}+\dfrac{\epsilon  \left(\pi ^2 k^2
   r_h^2-3 r^2\right)^2}{4 \pi ^5
   k^4 r^2 r_h^2} \sin \left(2 \pi  k\dfrac{ r_h}{r}\right) \notag \\ 
&& +\dfrac{3 \epsilon  \left(2 \pi ^4 k^4 r_h^4-6 \pi ^2 k^2
   r^2 r_h^2+3 r^4\right) }{8 \pi ^6
   k^5 r r_h^3}\cos \left(2 \pi  k\dfrac{ r_h}{r}\right).
\eeq
The density decays like ${\cal O}(r^{-5})$ at large distances and is finite at $r=r_h$. 
Figure~\ref{fig:PertMasFunc_DertPer} shows the mass function and associated energy density (inset) related to a potential modification in the form \eqref{eq:deltaV_sin} with $\epsilon=10^{-5}$ and $k=10$. Even though the mass function $m(r)$ is well defined in the outer domain $r\in[r_h,\infty)$, the energy density oscillates around the vacuum and assumes negative values. Therefore, these modifications are not realistic at a classic level either.

\subsection{The continued-fraction parametrisation}

\subsubsection{The spacetime and some of its properties}
%
\begin{table}[ht!]
\setlength{\tabcolsep}{9pt} 
\renewcommand{\arraystretch}{0.7} 
\begin{tabular}{|l||l||l|l|l|l|} 
\hline
Model      & $4\pi r_{h}$ T  & $a_0$ & $a_1$     & $a_2$   & $a_3$ \\
  \hline\hline
  BH0     & $1$       &    $0$       & $0$           & $-$     & $-$   \\\hline
  BH1     & $1.0001$  &    $0$       & $10^{-4}$     & $-10^3$ & $1001$  \\ \hline
  BH2     & $1.5$     &    $0$       & $0.5$         & $10^2$  & $0$    \\ \hline
  BH3     & $1$       &    $10^{-4}$ & $10^{-4}$     & $-2$    & $10^3$    \\
  \hline
\end{tabular}
\caption{Parameterized BH spacetimes studied in this work. All other parameters (such as $b_0, b_1, a_4$) are set to zero, while $\epsilon=a_0$. The vacuum Schwarzschild spacetime corresponds to model BH0. Models BH1 and BH2 follow Ref.~\cite{Konoplya:2022pbc}, with BH2 spectrally unstable. BH3 assumes conditions~\eqref{eq:cond_a_b} to make it physically interesting, but is still spectrally unstable. }
\label{tab:BH_models}
\end{table}
Finally, we discuss one other instance of reported high-overtone instability. This concerns a generic parameterization of an axi-symmetric spacetime, which in the non-rotating limit reduces to parameterizing the metric functions \eqref{eq:metric} via~\cite{Rezzolla:2014mua,Konoplya:2016jvv}
\beq
\label{eq:metric_functions_KRZ}
a &=& \left( 1- \dfrac{r_h}{r} \right)A\,,\qquad b = \dfrac{\left( 1- \dfrac{r_h}{r} \right)A}{B^2}\,.
\eeq
The functions $A$ and $B$ are expressed in terms of a compact radial coordinate centered at the horizon 
$
x=1-{r_h}/{r}
$,
as
\beq
A&=& 1 - \epsilon(1-x) + (a_0 - \epsilon)(1-x)^2 + \tilde A (1-x)^3 \\
B&=& 1 + b_0(1-x) + \tilde B(1-x)^2.
\eeq
The functions $\tilde A(x)$ and $\tilde B(x)$ are expressed via continued fractions 
\beq
\tilde A=\dfrac{a_1}{1 + \dfrac{a_2 x}{1 + \dfrac{a_3 x}{1+\cdots}}}, \quad
\tilde B =\dfrac{b_1}{1 + \dfrac{b_2 x}{1 + \dfrac{b_3 x}{1+\cdots}}} \, .
\eeq
These definitions ensure that the BH horizon is located at $r=r_h$. The Hawking temperature reads
\be
T=\dfrac{(1+a_0 +a_1 -2 \epsilon)}{4\pi\, r_h (1+b_0+b_1)}\,.\label{eq:Hawking_pars}
\ee

The matter content is anisotropic, with a stress tensor of the form ${\rm diag}(-\rho,p_r,p_t,p_t)$. At the horizon, 
\beq
8 \pi r_h^2 \rho = - 8 \pi r_h^2 p_r = 1 - \dfrac{1+a_0+a_1-2 \epsilon}{\left(1+b_0+b_1\right)^2}\,.\label{eq:density_KRZ}
\eeq
The tangential pressure $p_t$ looks slightly more cumbersome and we refrain from writing it here.

\subsubsection{Spectral properties}
%
\begin{figure*}[th!]
\includegraphics[width=0.49\linewidth
]{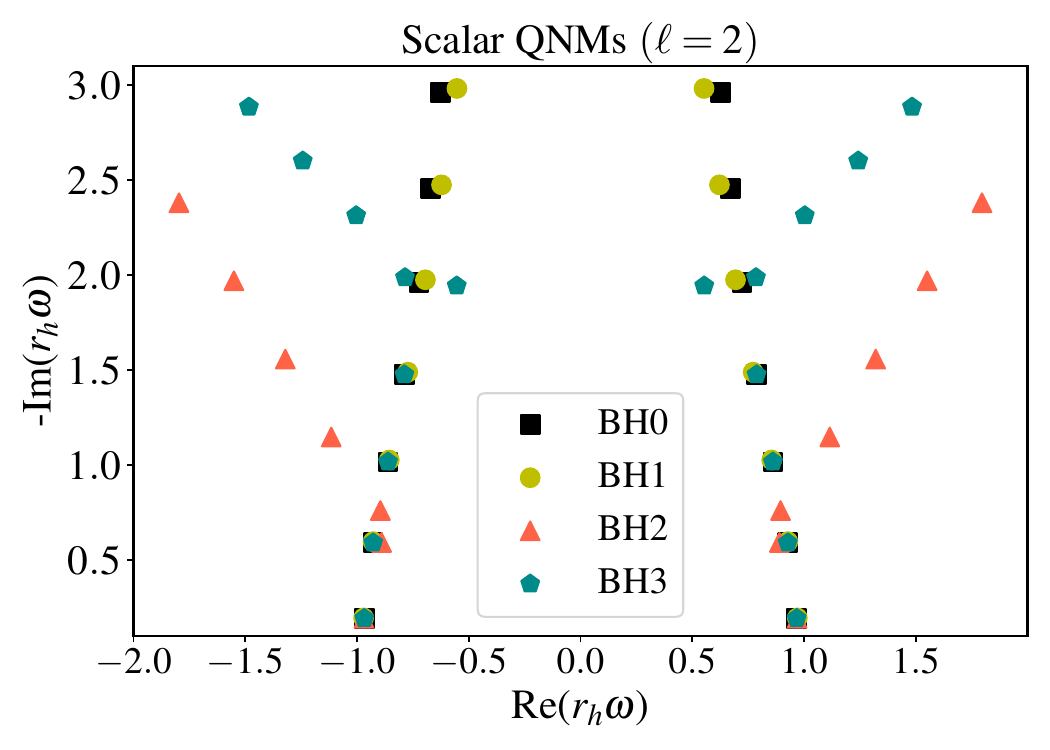}	
\includegraphics[width=0.49\linewidth
]{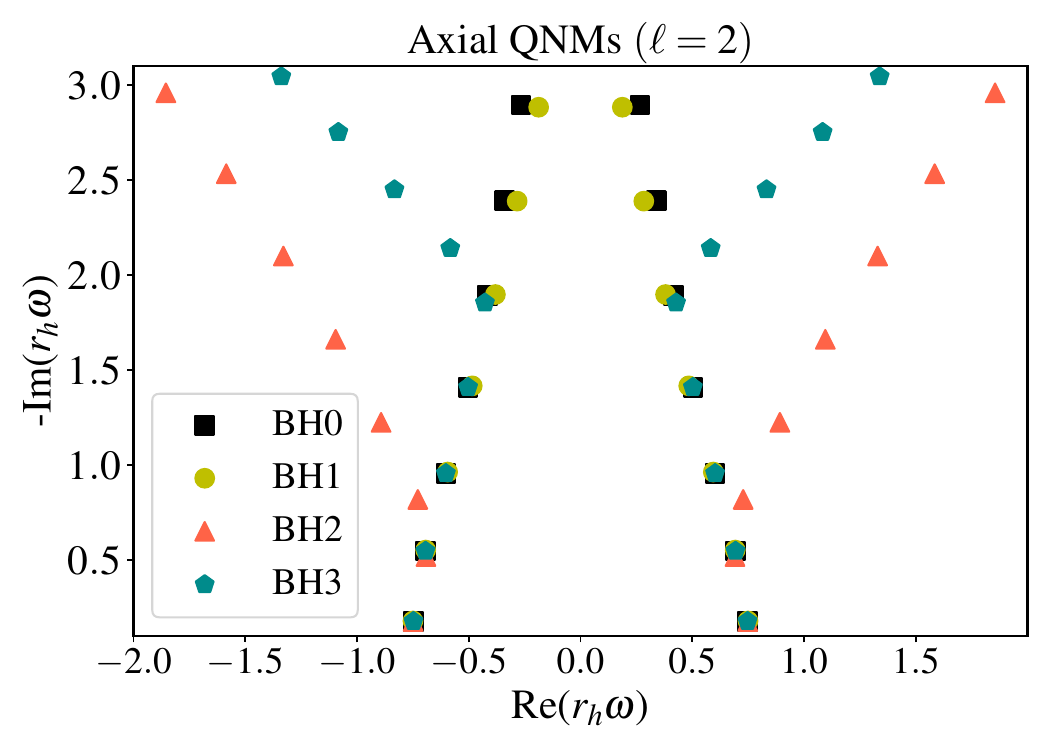}	
\caption{{\bf Left Panel:} Scalar QNMs for the BH models with parameters given in Table \ref{tab:BH_models}. Overtone instability is triggered in models BH2 and BH3, in particular with the appearance of a new $n=2$ overtone for BH2 not reported in ref.~\cite{Konoplya:2022pbc}. {\bf Right Panel:} Axial QNMs for the BH models with parameters given in Table \ref{tab:BH_models}. As in the scalar sector, overtone instability is triggered for the models BH2 and BH3.}\label{fig:QNMs_KRZ}
\end{figure*}
We focus on three specific models, two of those explored in ref.~\cite{Konoplya:2022pbc}, summarized in Table~\ref{tab:BH_models}. In particular, model BH0 corresponds to the Schwarzschild spacetime, whereas BH2 is spectrally unstable as we discuss shortly. We list the Hawking temperature~\eqref{eq:Hawking_pars} of each of these spacetimes.

We first reproduce the scalar QNMs reported in the original work~\cite{Konoplya:2022pbc}. Figure~\ref{fig:QNMs_KRZ} summarizes our results for quadrupolar $\ell=2$ modes. We employ the hyperboloidal framework, with Chebyshev spectral methods to calculate the QNMs~\cite{Jaramillo:2020tuu,PanossoMacedo:2023qzp}, and ensure that the numerical resolution provides enough accuracy. We reproduce scalar QNMs exactly for all models BH0, BH1, BH2, confirming the correct implementation of the hyperboloidal framework\footnote{Note, however, that we find a new overtone $n=2$, absent in Ref.~\cite{Konoplya:2022pbc}. The real part of this overtone is very close to the previous one: ${\rm Re}(r_h \omega_1) = 0.890084$, and ${\rm Re}(r_h \omega_2) = 0.895286$. Our method computes the QNM as the eigenvalues of the  numerically discretised hyperboloidal operator, and therefore, it calculates all overtones at once. Alternative methods, such as Leaver's continued fraction, are based on a root search algorithm, for which an initial seed relatively close to a given overtone $n$ is required. Since the overtone $n=1$ and $n=2$ are rather close to each other, it is not unexpected that the root search may converges to $n=1$ and misses the new $n=2$, unless a fine tuned initial seed is specified.}.

We have extended the analysis to axial gravitational fluctuations, in Fig.~\ref{fig:QNMs_KRZ}. Overtone instabilities are present in model BH2 and BH3, clearly seen in Fig.~\ref{fig:QNMs_KRZ}. An inspection of \eqref{eq:density_KRZ} shows that $8\pi r_h^2\rho$ goes from $-10^{-4}$ for BH1 to $-0.5$ to model BH2, hardly a small number, and {\it negative}.

\subsubsection{A realistic model}
As might be anticipated from the discussion on charged BHs, there are physically relevant situations for which the spectrum is unstable. To search for these we consider the following, somewhat stringent, assumptions: 
\beq
a_0 = a_1 = \epsilon,\quad a_2=-2, \quad b_0=b_1=0\,.\label{eq:cond_a_b}
\eeq
These conditions ensure that there is no modification to the BH temperature, $4\pi \, r_h T = 1$,
that the density, tangential and radial pressure vanish at the horizon, that the geometry agrees with observational constraints on PN parameters, and that $\displaystyle \lim_{r\rightarrow \infty} m(r) = M_{\rm ADM}$.

With this setup, the small parameter $\epsilon$ provides only an overall re-scale on the amplitudes of $\rho$, $p_r$ and $p_t$. We complement conditions above with $a_4=0$, reducing the parameter space to the quantities $(\epsilon, a_3)$. For this model, which we dub BH3 in Table~\ref{tab:BH_models} and Fig.~\ref{fig:QNMs_KRZ}, the density distribution reads 
\beq
& 8\pi r_h^2\, \rho = \epsilon  \left(1 -\dfrac{r_h}{r}\right)^2 \left(\dfrac{r_h}{r}\right)^4 \times \\
&\times  \dfrac{  2 \left(a_3^2+a_3-2\right) r r_h+(a_3-1)^2 r^2-3 (a_3-2) a_3 r_h^2}{\bigg(r(1-a_3)+(a_3-2) r_h\bigg)^2}. \notag
\eeq
To ensure $\rho(r)\geq 0$ for $r>r_h$, we take $a_3\geq1$. 

In particular, the density profile has a peak $\left.8 \pi r_h^2 \,  \rho\right|_{\rm peak} =\, 16\epsilon/243$ exactly at the photon sphere $r=3 r_h/2$ when $a_3=1$. As $a_3\rightarrow \infty$, the peak moves slightly towards the horizon up until $r=(\sqrt{97}-5 ) r_h/4\approx 1.21 \, r_h$, where it takes the value $8 \pi r_h^2 \, \rho =(512 (17 -\sqrt{97}) \epsilon)/(\sqrt{97}-5)^6= 2.82\times 10^{-1}\, \epsilon$. Besides, we also observe that corresponding potential for axial perturbations do not develop any further peak, and that its relative difference with respective to the Regge-Wheeler potential has a maximum of order $\sim 10^{-1}\, \epsilon$ around $r\sim 2 r_h$.

Summarizing, BH3 seems to have all ingredients of a ``reasonable'' spacetime. Nevertheless, Fig.~\ref{fig:QNMs_KRZ} shows clearly that it also turns the BH spectrum unstable, even when it should be only a minor modification away from a vacuum, Schwarzschild BH. Notice that the spacetime was constructed to be reasonable and forced to obey a number of unnecessary restrictions (coefficients after $a_4$ are set to zero, etc). Thus, we believe that spectral instabilities of high overtones are a generic effect, and the arguments underlying the asymptotic analysis support this claim~\cite{Motl:2003cd,Andersson:2003fh,Natario:2004jd,Berti:2009kk}.

To study the role played by the parameters $a_3$ and $\epsilon$ in the instability, we define the critical mode $n_c$ as the overtone for which the relativity difference to its Schwarzschild counterpart is larger than the small parameter $\epsilon$. Specifically, the critical mode $n_c$ is defined via the relation $\delta \omega_n > \epsilon$ for $n>n_c$, with
\be
\delta \omega_n = \dfrac{\left|\omega_n^{\rm BH3} - \omega_n^{\rm BH0}   \right|}{\left|\omega_n^{\rm BH0}\right|}.
\ee
This definition captures the notion of QNM instability, and the model BH3 allows us to assess the two main important features underlying the phenomena of QNM instability. While $\epsilon$ controls the overall scale of how intense the perturbation is, $a_3$ impacts the opening of the QNM branches. Thus, the bigger either values, the lower the value $n_c$ for the critical mode. For instance, when $\epsilon = 10^{-4}$, we observe $n_c = 6$ for $a_3 \sim 10$, $n_c=5$ for $a_3 \sim 25$, $n_c = 4$ for $50 \lesssim a_3 \lesssim 250$ and $n_c = 3$ for $a_3 > 500$. Increasing $\epsilon$ allows one to obtain lower values of $n_c$, but their dependence on $a_3$ remains unaltered. 

\section{Discussion}
The purpose of this work is to shed light on spectral instabilities in BH spacetimes. One first important point is that some scenarios used to study this issue are not realistic. Matter is either extreme, or the spacetime is extreme and not a fluctuation away from the intended background.

More importantly, we argue that instabilities come in two broad classes \cite{Jaramillo:2020tuu} and both can be triggered with reasonable physics. Fundamental-mode instabilities, of direct interest to current detectors, involve a new scale in the problem. For many compelling scenarios these de-stabilize the spectra. We argued that these affect the signal only at very late times when the signal is already weak. However, a perturbation theory approach indicates that the changes in the frequency of the modes behave as $\delta \omega \sim e^{2L\omega_I}$, showing therefore an exponential dependence on the overtone number $n$ and on the new lengthscale $L$. This argument indicates an urgent need to understand mode content in time-domain waveforms.
The above concerns ``soft changes,'' but near-horizon structure or ``hard'' conditions somewhere in the spacetime is known to result in ``echoes'' of significant amplitudes.

The second class of fluctuations has no obvious new physical scale associated with it, and changes the asymptotic overtone structure of the spectrum. As we remarked, a result well-known for years (surprisingly not mentioned in this context in the literature) concerns slightly charged BHs. But there are other examples. The analysis and conclusions of Refs.~\cite{Motl:2003cd,Andersson:2003fh,Natario:2004jd,Berti:2009kk} suggests that the asymptotic structure of the spectrum relates to the behavior of the potential close to the {\it singularity}, a remarkable statement bearing in mind that these are cloaked by horizons.
The impact of the large overtone structure on our understanding of BH spectroscopy is unknown. In other words, some theories might leave the first few modes unaffected, possibly producing a ringdown signal identical to that of GR at late times. We do not have a precise understanding of the impact of the large overtone structure on time domain waveforms, but first attempts indicate the overtone instability is measurable in ideal highly accurate setups~\cite{Jaramillo:2021tmt}. Thus, ours and all the recent results in the literature make this an important issue.

\acknowledgments
We are thankful to Gregorio Carullo, Thomas Spieksma and Jaime Redondo Yuste for a critical reading of the manuscript and for useful feedback.
We are thankful to the Yukawa Institute for Theoretical Physics, for warm hospitality.
We acknowledge support by VILLUM Foundation (grant no. VIL37766) and the DNRF Chair program (grant no. DNRF162) by the Danish National Research Foundation.
V.C.\ is a Villum Investigator and a DNRF Chair.  
V.C. acknowledges financial support provided under the European Union’s H2020 ERC Advanced Grant “Black holes: gravitational engines of discovery” grant agreement no. Gravitas–101052587. 
Views and opinions expressed are however those of the author only and do not necessarily reflect those of the European Union or the European Research Council. Neither the European Union nor the granting authority can be held responsible for them.
This project has received funding from the European Union's Horizon 2020 research and innovation programme under the Marie Sklodowska-Curie grant agreement No 101007855 and No 101131233.

\bibliography{reference.bib}

\end{document}